\documentclass[%
reprint,
superscriptaddress,
 amsmath,amssymb,
 aps,
]{revtex4-1}

\usepackage{graphicx}
\graphicspath{{./}}
\usepackage{epstopdf}
\usepackage{psfrag}
\usepackage{dcolumn}

\usepackage{bm}
\usepackage{color}
\definecolor{darkblue}{rgb}{0,0,.8}
\usepackage{hyperref}
\hypersetup{pdftex=true, colorlinks=true, breaklinks=true, linkcolor=darkblue, citecolor=darkblue, menucolor=darkblue, pagecolor=darkblue, urlcolor=blue}

\usepackage[inline]{enumitem}
\usepackage{gensymb}

\usepackage[english]{babel}
\usepackage{listings}
\begin{document}

\title{Dielectric properties of nano-confined water: a canonical thermopotentiostat approach}

\author{F. Dei\ss enbeck}
 \affiliation{Max-Planck-Institut f\"ur Eisenforschung GmbH,
Max-Planck-Stra{\ss}e 1,
40237 D\"usseldorf, Germany
}

\author{C. Freysoldt}
 \affiliation{Max-Planck-Institut f\"ur Eisenforschung GmbH,
Max-Planck-Stra{\ss}e 1,
40237 D\"usseldorf, Germany
}

\author{M. Todorova}
 \affiliation{Max-Planck-Institut f\"ur Eisenforschung GmbH,
Max-Planck-Stra{\ss}e 1,
40237 D\"usseldorf, Germany
}

\author{J. Neugebauer}
 \affiliation{Max-Planck-Institut f\"ur Eisenforschung GmbH,
Max-Planck-Stra{\ss}e 1,
40237 D\"usseldorf, Germany
}

\author{S. Wippermann}
 \email{wippermann@mpie.de}
\affiliation{Max-Planck-Institut f\"ur Eisenforschung GmbH,
Max-Planck-Stra{\ss}e 1,
40237 D\"usseldorf, Germany
}



\date{\today}

\begin{abstract}
\renewcommand*{\figurename}{Abb.}

We introduce a novel approach to sample the canonical ensemble at constant temperature and applied electric potential. Our approach can be straightforwardly implemented into any density-functional theory code. Using thermopotentiostat molecular dynamics simulations allows us to compute the dielectric constant of nano-confined water without any assumptions for the dielectric volume. Compared to the commonly used approach of calculating dielectric properties from polarization fluctuations, our thermopotentiostat technique reduces the required computational time by two orders of magnitude.

\end{abstract}

\maketitle

Molecular dynamics (MD) has become an indispensable tool to efficiently simulate the behaviour of a wide range of systems. Experiments, however, are often performed using basic thermodynamic variables that are different from the ones easily accessible in simulations. The need for constant temperature as opposed to the much simpler constant energy simulations is widely recognized as one of the most important examples of this kind. Consequently, significant effort has been directed at developing thermostats \cite{woodcock,andersen,berendsen,nose,hoover,nhchain,schneider,langevin,bussi1,bussi2,braun} with the dual purpose of (i) efficiently sampling the canonical ensemble and (ii) enabling direct control of the temperature. With the advent of robust techniques to apply electric fields in density-functional calculations \cite{resta,stengel,stengel2,lozovoi,tavernelli,otani,esm,greens,zhang,sayer,yang,ashton,surendralal,magnussen}, there has been continuous interest to use MD simulations to study electrically triggered processes involving electron transfer reactions, such as electrochemical reactions, field desorption, and quantum transport. For this purpose, it is necessary to conceive a \emph{potentiostat}, in analogy to the theory of thermostats, in order to incorporate the electric potential as a thermodynamic degree of freedom.

\begin{figure}[t]
  \centering
    \includegraphics[width=0.48\textwidth]{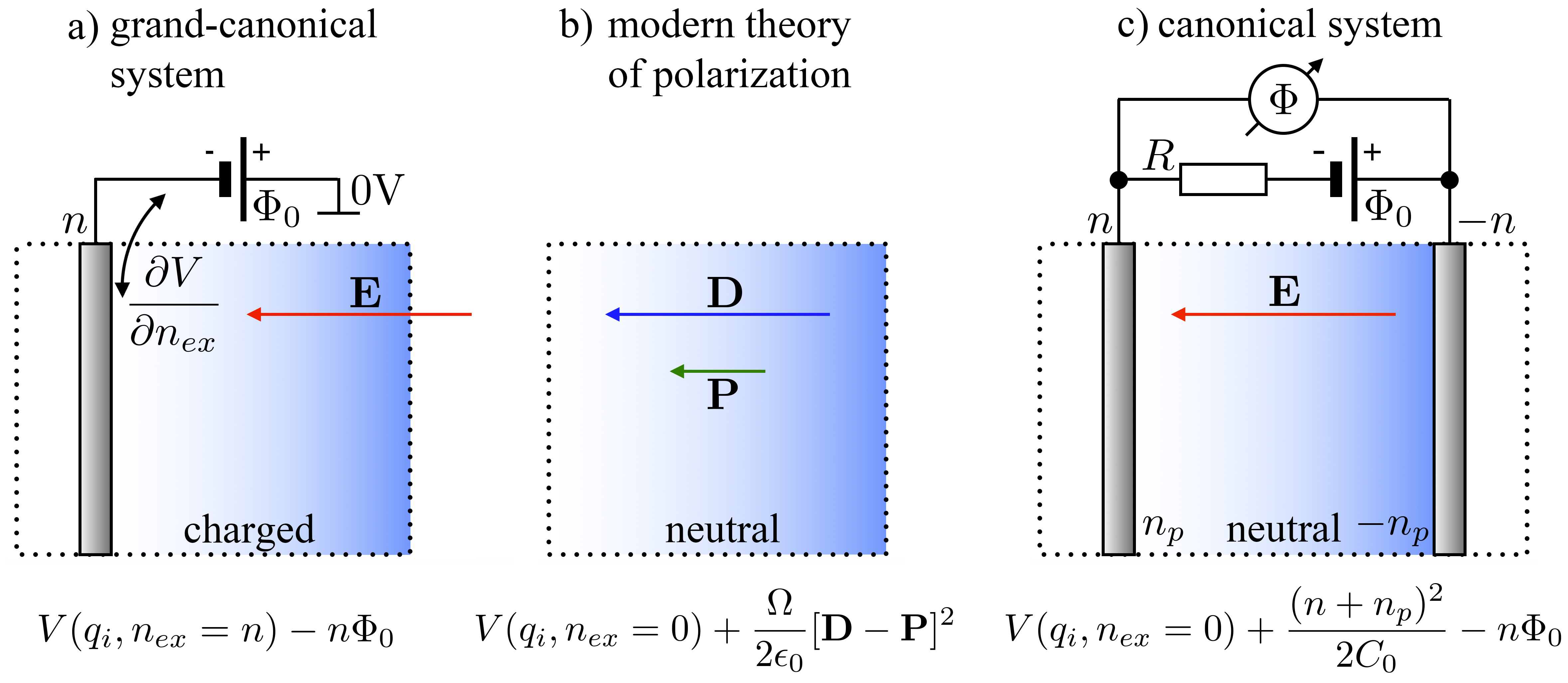}
\caption{\label{ens} Various computational setups used to include the effect of an applied electric field. \textbf{a)} The simulation cell contains a single charged electrode. It is grand-canonical and has an excess charge $n_{ex} = n$ \cite{lozovoi,otani}. \textbf{b)} In the modern theory of polarization (MTP), either the electric field $\mathbf E$ or the displacement field $\mathbf D$ is used as the basic variable \cite{stengel2}. The simulation system is always charge-neutral. \textbf{c)} The simulation cell contains two electrodes with charge $n$ and $-n$. In analogy to the MTP, it is canonical and charge-neutral.}
\end{figure}

The first approach of this kind was pioneered by Bonnet \emph{et al.} \cite{otani}. They chose a computational setup that is grand-canonical with respect to the electronic charge, cf. Fig. \ref{ens}a. The system is then coupled to an external potentiostat, where the charge is treated as an extended spatial coordinate. Analogous to how the mechanical momenta are obtained from the derivatives of the Hamiltonian with respect to the spatial coordinates, a fictitious momentum of the charge is obtained from the derivative of the total energy with respect to the charge. This fictitious momentum is then coupled to standard Nose-Hoover dynamics. The grand-canonical nature of their setup led Bonnet \emph{et al.} to recognize that ''\emph{[...] in its present formulation, a requirement for implementing the potentiostat scheme is the existence of an energy function $\mathcal{E}(r_i,n_e)$ that is differentiable with respect to the total electronic charge. This implies the ability to treat non-integer numbers of electrons and, in general, non-neutral systems.}`` \cite{otani}. Unfortunately, in the context of density-functional calculations the total energy as a function of the number of electrons is a notoriously difficult quantity to compute. Furthermore, the electronic charge is a single degree of freedom. Yet, thermostating single degrees of freedom by the Nose-Hoover method requires a chain of Nose-Hoover thermostats \cite{nhchain}. Both requirements are serious obstacles in implementing this potentiostat concept in existing density-functional theory (DFT) codes. In fact, none of the commonly used DFT codes \cite{sandia} has a potentiostat scheme that provides the opportunity to study electrochemical systems with molecular dynamics and explicit water.

In the present study we show that the origin of these difficulties lies in the formally equivalent construction of the potentiostat to a thermostat: existing approaches consider a transfer of energy (thermostat) or charge (potentiostat) from the DFT cell to an external reservoir. While such a grand-canonical coupling can be straightforwardly implemented for an energy exchange this task is severely harder for a charge transfer since non-neutral DFT systems have to be considered. Moreover, there is also a fundamental difference between thermostats and potentiostats. For a thermostat, the system is considered to be embedded in an external temperature bath. In the physical realization of such a setup, heat transfer occurs only at the boundaries between the system and the temperature bath. Therefore, in simulations, it is common to model the simulation cell, the thermostat, and the energy exchange between them as separate entities. For a potentiostat, however, the energy exchange with the voltage source is mediated by the electric field. Since the electric field permeates the system, energy exchange occurs throughout the whole system and not just at the boundaries.

Thus, the electric field plays the same conceptual role for a potentiostat as the temperature bath for a thermostat. Yet, the electric field is an integral part of the system. In contrast to a thermostat where the bath is external, only the \emph{control mechanism} of a potentiostat is external, but not the bath. We therefore propose to construct a potentiostat using the electric field as the control parameter. As will be derived in the following, this allows us to remove the need for treating charged systems. The actual implementation requires only quantities that are readily accessible in standard DFT codes, which makes it easy to integrate this potentiostat into existing electronic structure codes.

We start our derivation from the Hamiltonian given by Bonnet \emph{et al.} \cite{otani} for the grand-canonical potentiostat scheme sketched in Fig. \ref{ens}a:
\begin{equation}
\mathcal H = K(p_i) + V(q_i,n_{ex} = n) - n\Phi_0.
\end{equation}
The explicit variables to describe atoms exposed to an external electric field are the atoms' positions $q_i$, their momenta $p_i$ and the kinetic energy of the system $K(p_i)$. The single electrode present in the grand-canonical setup carries a charge $n$ and is included in the simulation cell. A direct consequence of having a single charged electrode is that the system is not charge neutral but has an excess charge $n_{ex}=n$. The potential energy $V(q_i, n_{ex})$, which describes the interatomic interaction, thus explicitly depends on the (non-zero) excess charge. We note that the potential energy $V(q_i, n_{ex} = 0)$ is simply the Born-Oppenheimer surface as computed in any DFT code. The term $-n\Phi_0$ accounts for the potential of the external charge reservoir. To derive an equation of motion for the potentiostat from this Hamiltonian requires an explicit expression for the derivative of the charged system's energy $V(q_i,n_{ex})$ with respect to the excess charge $n_{ex}$ \cite{otani}.

A key concept of our proposed approach is to remove the need to compute $V(q_i,n_{ex})$ explicitly. As an intermediate step to achieve this aim, we consider the system shown in Fig. \ref{ens}b, which is based on concepts of the modern theory of polarization \cite{resta}. Here, the simulation cell contains the dielectric displacement field $\mathbf D$ created by moving a charge $n$ from the left to the right boundary. The charge itself is outside the simulation cell. The total of the left and the right charge is zero, i.e., the inner region of the cell remains charge neutral. Without loss of generality, $V$ can then be partitioned into the regular interatomic potential energy at zero excess charge $V(q_i, n_{ex} = 0)$ and the electric energy $E(q_i,n)$ due to the presence of the $\mathbf D$-field. The electric energy $E$ is given by $E = \frac{\Omega}{2\epsilon_0} [\mathbf{D} - \mathbf{P}]^2$ \cite{stengel2}, where $\mathbf{P}$ is the polarization density. Thus, the difficult task of calculating $V(q_i,n_{ex}=n)$ for the charged system is substituted by the much easier calculation of $V(q_i,n_{ex}=0) + E(q_i,n)$ for a corresponding neutral system.

In a final step we extend this idea to a dielectric placed between two electrodes. The electrodes are connected to a voltage source with potential difference $\Phi_0$ and an internal resistance $R$, cf. Fig. \ref{ens}c. The corresponding Hamiltonian is \cite{SI}:
\begin{equation}
\mathcal H = K(p_i) + V(q_i, n_{ex} = 0) + \frac{(n + n_p)^2}{2 C_0} - n\Phi_0. \label{Ham}
\end{equation}
Here, we explicitly include the two electrodes with charge $n$ and $-n$ that create the displacement field $\mathbf D$. $C_0$ is the bare capacitance of the electrodes in vacuum without the dielectric and $n_p = \Phi/C_0 - n$ is the polarization bound charge at the left electrode surface due to the polarization of the dielectric \cite{landau}, where $\Phi$ is the instantaneous voltage measured directly across the electrodes. Note that the bound charge $n_p$ is an implicit function of $q_i$ and $n$. The Hamiltonian of the canonical potentiostat (Eq. \ref{Ham}) allows us to obtain $\partial \mathcal H / \partial n$ analytically:
\begin{equation}
\frac{\partial \mathcal H}{\partial n} = \frac{n + n_p}{C_0} - \Phi_0 = \Phi - \Phi_0. \label{displ}
\end{equation}
The reason why we are now able to obtain an analytical expression is that in the canonical system, $n$ is no longer the net charge $n_{ex}$ of the system exchanged with an external bath as in the grand-canonical potentiostat approach. Rather, it is based on a \emph{charge transfer} from one electrode to the other which leaves the total charge of the system unchanged. Hence, $\partial \mathcal H / \partial n$ is determined by the infinitesimal amount of energy $\partial \mathcal H$ required to transfer an infinitesimal amount of charge $\partial n$ between the electrodes. Thereby, Eq. \ref{displ} connects the derivative $\partial \mathcal H / \partial n$, that has to be computed for the microscopic quantum-mechanical system, to an external macroscopic quantity: the instantaneous voltage $\Phi$. Since our total system is charge neutral, $\Phi$ is determined directly by the total dipole moment of the charge distribution within our simulation cell. This quantity is readily accessible in DFT codes. In the following, we will use these properties of Eq. \ref{displ} to construct an equation of motion for the voltage $\Phi$ and, subsequently, for the electrode charge $n$, the control parameter of our potentiostat.

A thermostated system evolves under the influence of its internal, energy-conserving Hamiltonian and extra forces that drive the exchange of energy with the thermal bath. Similarly, a potentiostat scheme requires a force-like term that drives the necessary changes of the electrode charge to keep the average potential constant. To obtain this force we recast Eq. \ref{displ} in differential form:
\begin{eqnarray}
d\Phi &=& \frac{1}{C_0} dn_p + f dt \label{dphi}\\
f dt &=& \frac{1}{C_0} dn \label{fdt1}
\end{eqnarray}
The first term in Eq. \ref{dphi} expresses how the potential $\Phi$ will evolve under the system's internal, energy-conserving Hamiltonian. The second term, $f dt$, is the extra force-like term that couples to the electrode charge. This extra-force term controls on the one hand the constancy of the (average) potential. On the other hand, it has to mimic the statistical fluctuations that occur in a finite system. These two aspects are balanced by the fluctuation-dissipation theorem \cite{callen} and ensure that the system stays in the NT$\Phi$ ensemble.

Changing the potential of a capacitive system is always a dissipative process, i.e., only adapting the electrode charge to control the voltage would drain energy from the system. To avoid this energy drain, any dissipation must be accompanied by a corresponding fluctuation that returns the removed energy.  In other words, the applied electric field itself must have a finite temperature: the energy dissipated by the potential control mechanism must equal on average the energy gained from thermal potential fluctuations. For the electrical circuit shown in Fig. \ref{ens}c, Johnson \cite{johnson} and Nyquist \cite{nyquist} derived already in 1928 the relation between fluctuation and dissipation. This relation is now known as a specific case of the fluctuation-dissipation theorem (FDT), and determines the variance of the potential fluctuations as well as its distribution in frequency space. Using Ohm's Law and Kirchhoff's 2nd Law, the current through the circuit shown in Fig. \ref{ens}c is $dn = - (\Phi - \Phi_0) R^{-1} dt$. In conjunction with the FDT, we can then express the potentiostat force $f dt$ as a stochastic differential equation (SDE) \cite{SI}:
\begin{eqnarray}
f dt &=& \underbrace{-\frac{1}{\tau_{\Phi}} (\Phi - \Phi_0)\ dt}_{dissipation} + \underbrace{\sqrt{\frac{2}{\tau_{\Phi}} \frac{k_B T}{C_0}}\ dW_t}_{fluctuation}. \label{fdt2}
\end{eqnarray}
$\tau_{\Phi} := (RC_0)^{-1}$ and $dW_t$ are the relaxation time constant of the potentiostat and a stochastic noise term given by a Wiener process, respectively. The deterministic dissipation term in Eq. \ref{fdt2} is equivalent to Ohm's Law. The stochastic term provides the thermal fluctuations. Together, they satisfy the FDT exactly, also in frequency space \cite{SI}. Far from equilibrium, the deterministic part in Eq. \ref{fdt2} dominates and drives the system towards the target potential with a relaxation time of $\tau_{\Phi}$. Close to the target potential, the deterministic term becomes small and the stochastic term takes over to ensure that in thermodynamic equilibrium the canonical ensemble is correctly sampled.

The SDE Eq. \ref{fdt2} can be solved by employing It\=o calculus \cite{gardiner}. Using this calculus and Eq. \ref{fdt1}, Eq. \ref{fdt2} can be integrated analytically to a finite time step \cite{SI}
\begin{eqnarray}
n(t + \Delta t) = n(t) &-& C_0 \left(\Phi(t) - \Phi_0 \right) \left(1-e^{-\frac{\Delta t}{\tau_{\Phi}}} \right) \nonumber\\ &+& N \sqrt{k_B TC_0 \left(1 - e^{-\frac{2 \Delta t}{\tau_{\Phi}}}\right)}, \label{ft}
\end{eqnarray}
where $N$ is a Gaussian random number with $\langle N \rangle = 0$ and $\langle N^2 \rangle = 1$. Eq. \ref{ft} is the central result of our derivation: All quantities entering the right side can be either directly obtained from the DFT calculation or are known from the specific setup. Thus, the electrode charge $n$ can be directly computed at each MD time step. This central result allows us to include the potentiostating process in any standard discrete time step MD scheme since the only quantity that needs to be extracted from the MD is the potential $\Phi(t)$. The scheme can be used equally well to perform \emph{ab initio} or empirical potential MD.

\begin{figure}[t]
  \centering
    \includegraphics[width=0.42\textwidth]{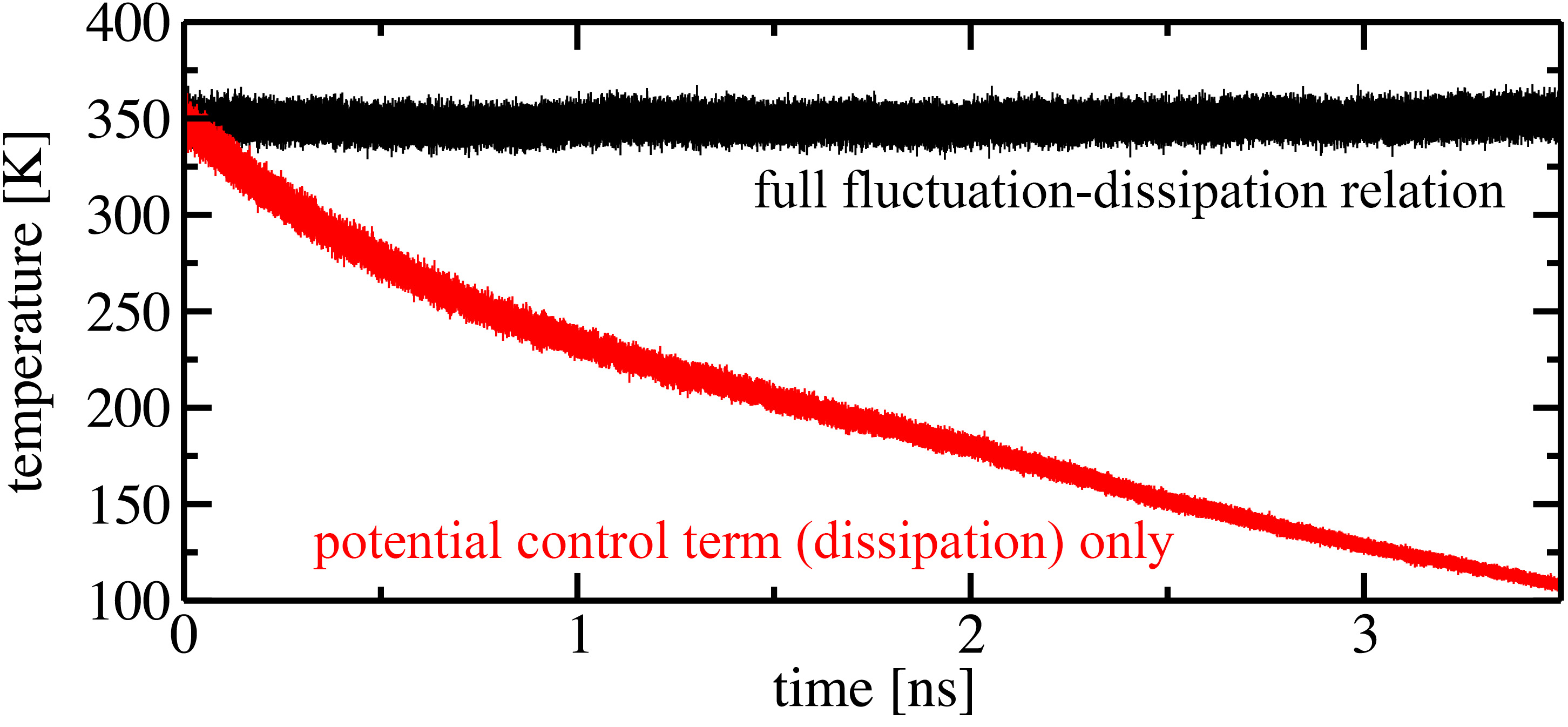}
\caption{\label{temp} Time evolution of the temperature for an NVE ensemble consisting of 1558 TIP3P water molecules, an electrode separation of $d = 8$ nm and potentiostated to $\Phi_0 = 0$ V. The electrode charge is adjusted, using only the dissipation (potential control) term in Eq. \ref{ft} (red line) or using the full Eq. \ref{ft} (black line).}
\end{figure}

\begin{figure*}[t]
  \centering
    \includegraphics[width=0.89\textwidth]{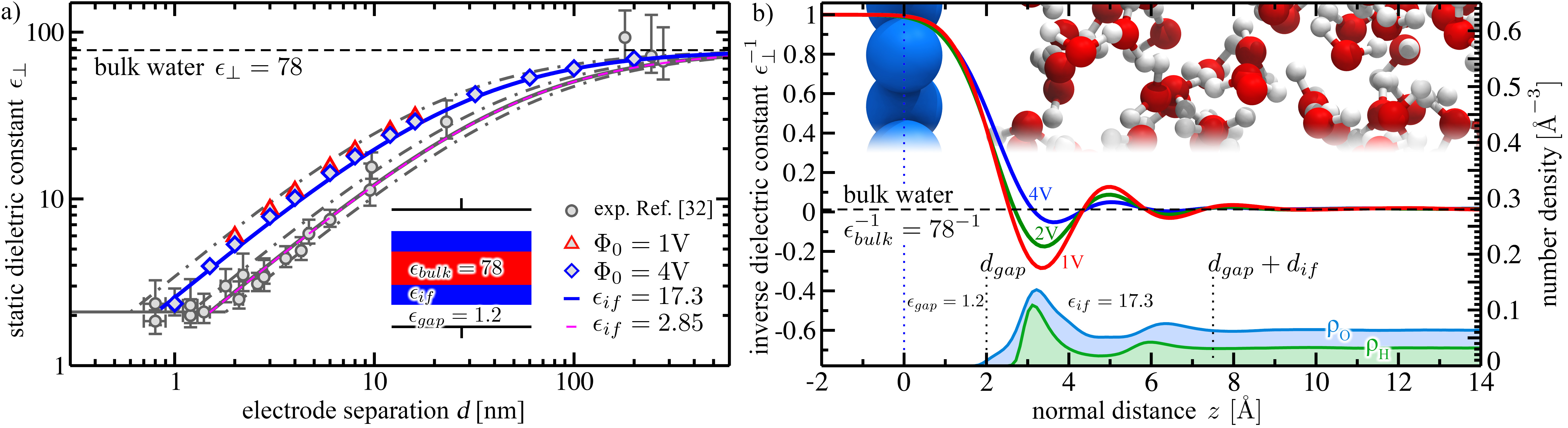}
\caption{\label{epsd} \textbf{a)} $\epsilon_{\perp}$ for TIP3P water as a function of electrode separation $d$, calculated using the NVT$\Phi$ ensemble. Experimental data reproduced with permission from Ref. \cite{fumagalli}. \textbf{b)} Local inverse dielectric profile and O/H number density profiles for $d = 8$ nm. The number density was computed at $\Phi_0 = 2$ V. The blue dotted line marks the position of the electrode.}
\end{figure*}

To validate our canonical thermopotentiostat scheme and to demonstrate its performance we consider a topic that had recently gained a lot of attention. Recent experiments \cite{fumagalli} showed that a water film confined to a few nm thickness changes its dielectric behaviour from the bulk dielectric constant of 80 down to 2. Thus, the presence of solid-water interfaces appears to modify the dielectric response of water from a highly polarizable medium, which is considered to be the origin of the unique solvation behaviour of water, down to a response that is close to the vacuum dielectric constant. Understanding and being able to qualitatively describe this mechanism is crucial since interfacial water is omnipresent.

Due to the relevance of this question in fields as diverse as electrochemistry, corrosion and electrocatalysis, several computational studies addressed dielectric properties of nano-confined water \cite{cui,loche,marx}. These studies used either the variance of the total dipole moment fluctuations per volume, using Kirkwood-Fr\"ohlich theory \cite{sharma}, or the theory of polarization fluctuations \cite{feller2}. They require as an input quantity the dielectric volume to compute the dielectric constant. However, the exact location of the boundary between the electrode and the dielectric is ill-defined in the presence of adsorbates, thermal motion of the electrode surface or in the context of explicit electronic structure calculations. For this reason, past studies often reported only dipole fluctuations perpendicular to the electrode surface, but not the dielectric constant $\epsilon_{\perp}$ itself \cite{cui}.

Our thermopotentiostat MD allows us to address this issue directly, since our setup shown in Fig. \ref{ens}c exactly reproduces the experimental situation. In analogy to experiment, we compute the static dielectric constant as \mbox{$\epsilon_{\perp} := \langle C \rangle_{\Phi_0} / C_0$}. The calculation of the capacitances requires only the averaged charge $\langle n \rangle$ and potential $\langle \Phi \rangle$. Thus, similarly to the experiment, our approach does not require a definition of the dielectric volume.

Before studying in detail the thickness dependent properties of nano-confined water let us use this system as a model to test our thermopotentiostat scheme. We therefore perform classical MD simulations \cite{lammps} of liquid TIP3P water confined between two parallel electrodes. Numerical details are given in the supplementary material \cite{SI}. We first investigated the effect of our potential control scheme on the temperature of an NVE ensemble. Applying the potential control mechanism alone without the fluctuation term, i.e., the upper part of Eq. \ref{ft}, dissipates thermally induced fluctuations of the total dipole moment. Thus, in absence of an explicit thermostat, the NVE ensemble shows a constant loss of kinetic energy and the system cools down, cf. Fig. \ref{temp}. Without a thermostat, but using the full fluctuation-dissipation relation Eq. \ref{ft}, this spurious energy transfer disappears. The temperature correctly fluctuates around the target temperature $T = 350$ K set in Eq. \ref{ft}. This result clearly demonstrates that our potentiostat alone is not only able to control the potential, but also the temperature, even in the absence of an explicit thermostat.

We now discuss the thickness-dependence of the dielectric properties. In Fig. \ref{epsd}a, we compare our calculated static dielectric constant $\epsilon_{\perp}$ as a function of the water layer thickness to the experimental data from Ref. \cite{fumagalli}. Red triangles and blue squares denote data points obtained from our thermopotentiostat MD at $\Phi_0 = 1$ V and $\Phi_0 = 4$ V, respectively. Experimental data points measured by Fumagalli \emph{et al.} \cite{fumagalli} are shown as grey circles. Consistent with the measurements, our results display a pronounced decrease of $\epsilon_{\perp}$ compared to the static dielectric constant of bulk liquid water $\epsilon_{bulk}$ that persists for electrode separations exceeding 100 nm. Based on this level of agreement, we therefore expect our TIP3P water model to correctly capture the impact of an interface on the dielectric properties of water.

In order to understand the origin of the decreasing $\epsilon_{\perp}$ with decreasing $d$, we computed the local static dielectric constant $\epsilon_{\perp}(z)$ as a function of the normal distance $z$ to the electrode surface \cite{SI}. Fig. \ref{epsd}b shows the inverse dielectric profile $\epsilon_{\perp}^{-1}(z)$ for an electrode separation of $d = 8$ nm. At the position of the electrode surface $z$ = 0, $\epsilon_{\perp}^{-1}$ drops sharply and intersects the water bulk value at $\sim$3 \AA\ above the surface. With further increasing $z$, $\epsilon_{\perp}^{-1}$ assumes negative values for interfacial water and then approaches the bulk water value in an oscillatory fashion. At a normal distance of \mbox{$\sim$9 \AA}, the dielectric constant of bulk liquid water $\epsilon_{bulk}$ is recovered. The behaviour in the $\sim$9 \AA\ thick interface layer reflects the well-known density modulations of water close to interfaces \cite{bopp,cherepanov}, cf. Fig. \ref{epsd}b lower part.

We will now test whether the presence of the relatively thin layer of interfacial water with modified dielectric properties explains the observed huge decrease of the total static dielectric constant. Guided by the dielectric profile shown in Fig. \ref{epsd}b, we partition the dielectric profile into three regions: (i) a hydrophobic gap between electrode and surface with a thickness of $d_{gap} = 2$\AA, (ii) an interfacial water region consisting of the first two water layers with a thickness of $d_{if} = 5.5$\AA\ and (iii) the remaining approximately bulk-like region (Fig. \ref{epsd}b). The existence of a hydrophobic gap, which is clearly visible in our data, was also suggested by Niu \emph{et al.} \cite{niu} based on spectroscopic data. The effective dielectric constant of each region is obtained by integrating over the dielectric profile, yielding $\epsilon_{gap} = 1.2$ and $\epsilon_{if} = 17.3$ for the hydrophobic gap and interfacial water regions, respectively. We further assume that each region is approximately independent of $d$. The surrogate electrostatic model becomes then a simple plate capacitor with multiple dielectrics (inset in Fig. \ref{epsd}a). The total dielectric constant is given by the analytical expression $\epsilon_{\perp}(d) = d / [2d_{gap}/\epsilon_{gap} + 2d_{if}/\epsilon_{if} + (d-2(d_{gap} + d_{if}))/\epsilon_{bulk}]$. The solid blue line in Fig. \ref{epsd}a denotes $\epsilon_{\perp}(d)$ obtained by the surrogate model. Although here $\epsilon_{\perp}(d)$ was obtained from a single explicit calculation for an electrode separation of $d = 8$ nm, the solid blue line accurately reproduces all other data points obtained from explicit thermopotentiostat MD simulations at different separations $d$. These findings confirm that the local dielectric properties of water close to the interface are indeed responsible for the observed reduction of $\epsilon_{\perp}$ compared to $\epsilon_{bulk}$.

The calculation of dielectric profiles from polarization fluctuations \cite{feller2} requires hundreds of nanoseconds of statistical sampling, in practice enforcing the use of classical MD. In contrast, the stochastic canonical sampling of our thermopotentiostat MD in conjunction with finite electric field techniques turns out to be extremely efficient: our expression for $\epsilon_{\perp}$ \cite{SI} allows us to rely purely on thermodynamic averages rather than variances. Thus, the dielectric profiles shown in Fig. \ref{epsd}b converged within less than 4 ns, reducing the required computational time by more than two orders of magnitude. 

In conclusion, we devised a novel thermopotentiostat approach to sample the canonical ensemble at constant temperature and applied electric potential. Our approach (i) avoids the need to treat non-neutral simulation cells, (ii) requires only quantities that can be directly obtained from density-functional theory simulations and (iii) is straightforward to implement in any standard \emph{ab initio} molecular dynamics package. To demonstrate the performance of our approach we computed the thickness-dependent dielectric properties of nano-confined water. We showed that the presence of interfaces strongly modifies the dielectric constant of an interfacial water region. This region is spatially highly confined with a thickness of only $\sim$1 nm (roughly two water-layers). This thin region with modified dielectric properties is shown to fully capture the experimentally observed anisotropic dielectric response, persisting for distances exceeding 100 nm. The computational efficiency of our approach is improved by more than two orders of magnitude compared to previous ones. In conjunction with \emph{ab initio} MD, we expect our thermopotentiostat to open the door towards accurate and efficient simulations of equilibrium properties, such as interfacial dielectric constants, as well as processes, such as electron transfer and electrochemical reactions.

We thank L. Fumagalli for providing the raw experimental data from Ref. \cite{fumagalli} and D. Marx for discussions. FD and SW are supported by the German Federal Ministry of Education and Research (BMBF) within the NanoMatFutur programme, grant no. 13N12972. Funded by the Deutsche Forschungsgemeinschaft (DFG, German Science Foundation) under Germany's Excellence Strategy -- EXC 2033 -- project no. 390677874. Supercomputer time provided by NERSC Berkeley, project no. 35687, is gratefully acknowledged.

\end{document}


\tcbset{highlight math style={boxsep=5mm,colback=blue!30!red!30!white}}

\title{Supplementary Information for\\ {Dielectric properties of nano-confined water: a canonical thermopotentiostat approach}}

\author{F. Dei\ss enbeck}
 \affiliation{Max-Planck-Institut f\"ur Eisenforschung GmbH,
Max-Planck-Stra{\ss}e 1,
40237 D\"usseldorf, Germany
}

\author{C. Freysoldt}
 \affiliation{Max-Planck-Institut f\"ur Eisenforschung GmbH,
Max-Planck-Stra{\ss}e 1,
40237 D\"usseldorf, Germany
}

\author{M. Todorova}
 \affiliation{Max-Planck-Institut f\"ur Eisenforschung GmbH,
Max-Planck-Stra{\ss}e 1,
40237 D\"usseldorf, Germany
}

\author{J. Neugebauer}
 \affiliation{Max-Planck-Institut f\"ur Eisenforschung GmbH,
Max-Planck-Stra{\ss}e 1,
40237 D\"usseldorf, Germany
}

\author{S. Wippermann}
 \email{wippermann@mpie.de}
\affiliation{Max-Planck-Institut f\"ur Eisenforschung GmbH,
Max-Planck-Stra{\ss}e 1,
40237 D\"usseldorf, Germany
}

\maketitle

\renewcommand{\thefigure}{S\arabic{figure}}
\renewcommand{\theequation}{S\arabic{equation}}
\renewcommand{\thetable}{S\arabic{table}}

\section{Hamiltonian}

In the modern theory of polarization, the Hamiltonian at constant dielectric displacement $\mathbf{D}$ is given by \cite{stengel2}
\begin{equation}
\mathcal{H} = E_{KS} + \frac{\Omega}{2\epsilon_0}[\mathbf{D} - \mathbf{P}]^2, \label{HamDP}
\end{equation}
where $E_{KS}$ and $\mathbf{P}$ are the Kohn-Sham energy and polarization density, respectively. For a plate capacitor, according to Gauss's Law, the value $||\mathbf D||$ of the dielectric displacement field between the plates is equal to the charge density $\frac{n}{A}$ on the plates. Analogously, we introduce an effective bound charge density $\frac{n_p}{A}$ on the plates that is equal to the value $-||\mathbf P||$ of the polarization density between the plates. With the definition of the capacitance $C_0 = \epsilon_0 \frac{A}{d}$ and $\Omega = A d$ we rewrite the electrical energy $E$ as:
\begin{equation}
E = \frac{\Omega}{2\epsilon_0}[\mathbf{D} - \mathbf{P}]^2 = \frac{[n + n_p]^2}{2 C_0}.
\end{equation}

We note that this step is optional. Instead of deriving an equation of motion for $n$ or $\Phi$ as done in our letter, it is also possible to use Eq. \ref{HamDP} in order to obtain an equivalent equation of motion for $\mathbf D$ or $\mathbf E$ in an analogous way. In the following, we use the charge $n$ as the basic variable. This allows us to introduce the Hamiltonian:
\begin{equation}
\mathcal{H}(p_i,q_i,n) = \sum_i \frac{p_i^2}{2m_i} + V(q_i, n_{ex}=0) + \frac{[n+n_p]^2}{2C_0} - n \Phi_0. \tag{2} \label{Ham}
\end{equation}
The potential energy $V(q_i, n_{ex} = 0)$ at zero excess charge $n_{ex}$ is simply the Born-Oppenheimer total energy surface as computed in any DFT code. If an explicit counter electrode is used in the DFT calculation, e.g., as in Ref. \cite{surendralal}, or the generalized dipole correction \cite{ashton}, the Kohn-Sham energy already includes the contribution from the electric field, so that $E_{KS} = V(q_i, n_{ex}=0) + \frac{[n+n_p]^2}{2C_0}$. Alternatively, the contribution from the electric field can be evaluated explicitly using the modern theory of polarization \cite{stengel2}.

\section{Equations of Motion}

The complete, self-contained set of equations of motion that samples the canonical ensemble at constant potential $\Phi_0$ is given by:

\begin{eqnarray}
dp_i &=& \left( \frac{\partial V}{\partial q_i} + \frac{\partial E}{\partial q_i} \right) dt + g_i dt \label{dpi}\\
dq_i &=& \frac{1}{m_i}p_i dt \label{dqi}\\
dn &=& -\frac{1}{\tau_{\Phi}} C_0 (\Phi - \Phi_0) dt + \sqrt{\frac{2}{\tau_{\Phi}} k_B T C_0} dW_t. \label{dpe}
\end{eqnarray}

Eqs. \ref{dpi} and \ref{dqi} are the standard Hamiltonian equations of motion, integrated by any molecular dynamics code. The term $g_i$ is used to couple to an external thermostat, cf. section \textsc{Thermopotentiostat MD with an explicit thermostat}. In the following, we outline how to derive the single additional equation of motion for the charge dynamics Eq. \ref{dpe}.

In the main text, we took the derivative $\frac{d}{dt} \frac{\partial \mathcal H}{\partial n}$ in order to obtain Eqs. 4 and 5 as:
\begin{align}
d\Phi =& \ \frac{1}{C_0} dn_p + f dt \tag{4} \label{dphi1}\\
f dt =& \ \frac{1}{C_0} dn. \tag{5} \label{fdt1}
\end{align}

We then equate the current flowing through the capacitance $dn/dt$ to the current $-(\Phi-\Phi_0)/R$ flowing in reverse through the voltage source and its internal resistivity $R$ (Ohm's Law and Kirchhoff's 2nd Law):
\begin{eqnarray}
f dt &=& \underbrace{- \frac{1}{RC_0}(\Phi - \Phi_0)dt}_{dissipation} +  \underbrace{\tilde{\Phi}dW_t}_{fluctuation}. \label{fdt}
\end{eqnarray}
Here, we add a fluctuation term $\tilde{\Phi}$ and a stochastic noise term given by a Wiener process $dW_t$, respectively. This term must be introduced in order to satisfy the fluctuation-dissipation theorem (FDT) \cite{callen}. Note that $dW_t$ plays an analogous conceptual role as $dt$. It is an integration variable and represents a stochastic time step for integration. Integration is performed by Ito calculus \cite{gardiner}, see the \textsc{Practical Implementation} section.

The FDT \cite{callen} determines both the variance and the frequency spectrum of the fluctuations. Classically, per frequency interval $d\nu$, the variance of the fluctuating voltage at a resistance $R$ is given by $\sigma^2_U d\nu= 4 k_B T R d\nu$ \cite{johnson,nyquist}. In conjunction with the capacitance $C_0$, the resistance $R$ in our system forms an RC low pass. The variance of $\Phi$ is obtained by integrating the noise spectral density over the bandwidth of the RC low pass:
\begin{equation}
\sigma^2_{\Phi} = 4 k_B T R \frac{1}{2\pi} \int_0^{\infty} \frac{1}{1 + (\omega R C_0)^2} d\omega = \frac{k_B T}{C_0}. \label{noise1}
\end{equation}
Eq. \ref{noise1} allows us to construct a suitable fluctuation term $\tilde{\Phi}$. The energy loss in Eq. \ref{fdt} due to the dissipation by the resistance $R$ must be exactly equal, on average, to the energy gained through the fluctuations in $\tilde{\Phi}$, while the variance, and also the frequency spectrum of the fluctuations, are determined by the FDT. Thereby, the applied electric field itself is fluctuating and has a finite temperature. In analogy to the Langevin \cite{langevin} and BDP \cite{bussi1,bussi2} thermostats, we recognize that Eq. \ref{fdt} formally represents a stochastic differential equation known as the Ornstein-Uhlenbeck process
\begin{equation}
dx = -k x\ dt + \sqrt{D}\ dW_t. \label{OUP}
\end{equation}
The expectation value and variance of the Ornstein-Uhlenbeck process are given by \cite{gardiner}:
\begin{eqnarray}
\langle x(t)\rangle &=& x_0 e^{-kt}\\
\mathrm{var}[x(t)] &=& \left( \mathrm{var}[x_0] - \frac{D}{2k} \right) e^{-2kt} + \frac{D}{2k}.
\end{eqnarray}
With $\tau_{\Phi} := RC_0$, $k := \tau_{\Phi}^{-1}$ and $D/2k := \sigma_{\Phi}^2$ we write:
\begin{equation}
f dt = -\frac{1}{\tau_{\Phi}} (\Phi - \Phi_0) dt + \sqrt{\frac{2}{\tau_{\Phi}} \frac{k_B T}{C_0}} dW_t. \tag{6} \label{fdt2}
\end{equation}
Eq. \ref{fdt2} provides a correction potential to the Hamiltonian evolution of the potential at constant dielectric displacement (constant charge). The corresponding charge dynamics given by Eq. \ref{dpe} is obtained trivially by multiplying Eq. \ref{fdt2} with $C_0$.

\section{The Nos\'e-Hoover Potentiostat as a Limiting Case of our Thermopotentiostat}

Bonnet \emph{et al.} proposed an empirically motivated potentiostat, introducing a fictitious momentum $P_{n}$ for the potential and and electronic mass $M_{n}$. These are used to perform Nos\'e-Hoover dynamics on the charge $n$, with 
\begin{eqnarray}
\dot{n} &=& \frac{P_n}{M_n} \label{dotn}\\
dP_n &=& (\Phi - \Phi_0) dt + P_n d\xi. \label{dpn}
\end{eqnarray}
$P_n^2 / (2M_n)$ enters the Nos\'e-Hoover equation of motion for $\dot{\xi}$ as a fictitious kinetic energy. From our scheme, it now becomes apparent that these quantities permit a direct physical interpretation. With
\begin{eqnarray}
P_n &:=& - R n \\ 
M_n &:=& \frac{R^2n}{\Phi - \Phi_0},
\end{eqnarray}
Eq. \ref{dotn} is identical to Ohm's law. The fictitious kinetic energy $P_n^2 / (2M_n)$ then becomes the electrostatic potential energy \mbox{$n(\Phi-\Phi_0)/2$}. Inserting $P_n$ and $M_n$ into Eq. \ref{dpn} yields:
\begin{equation}
dn = -\frac{1}{\tau_{\Phi}} C_0 (\Phi - \Phi_0) dt -n d\xi. \label{pnh}
\end{equation}
Comparing Eqs. \ref{pnh} and \ref{dpe}, we see that the deterministic $d\xi$ in Bonnet \emph{et al.}'s potentiostat is substituted in our approach by a fully stochastic term in Eq. \ref{dpe}.

\section{Spectral Considerations}

\begin{figure}[b]
  \centering
    \includegraphics[width=0.4\textwidth]{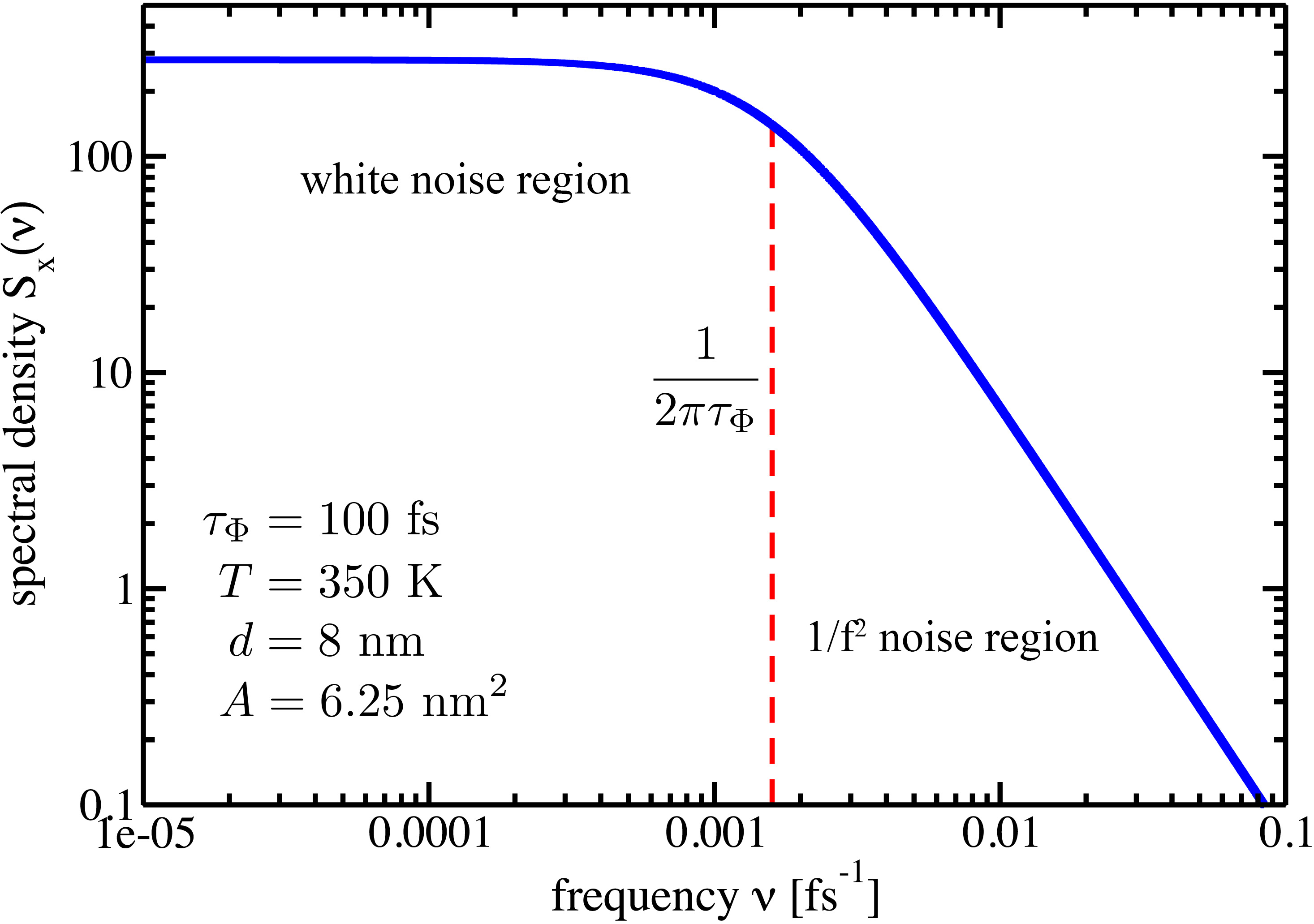}
\caption{\label{spect} Spectral density plot of the variance of the potential fluctuations.}
\end{figure}

The FDT determines both the variance and also the frequency spectrum of the fluctuations. In absence of a dielectric, the variance of the voltage fluctuations given by Eq. \ref{noise1} is distributed in frequency space according to the spectral density function \cite{gillespie1} of the ideal fully relaxed Ornstein-Uhlenbeck process
\begin{equation}
S_x(\nu) = \frac{2c\tau^2}{1 + (2\pi\tau\nu)^2},
\end{equation}
with $\tau = \tau_{\Phi}$ and $c = 2 k_B T / (\tau_{\Phi} C_0)$. In Fig. \ref{spect} we show the spectral density of the variance of the potential fluctuations for the $d = 8$ nm cell. Below the cutoff frequency $(2\pi \tau_{\Phi})^{-1}$, the slope of the spectral density is zero (white noise region), whereas it is -2 above ($1/f^2$ noise region), cf., e. g., Ref. \cite{gillespie1} for a detailed discussion. Note that the fluctuations are always independent of the dielectric properties of the system under consideration. Instead, the dielectric properties determine how the system responds to the excitation provided by the fluctuations.

Comparing the expressions for the Nose-Hoover potentiostat and our thermopotentiostat
\begin{eqnarray}
dn &=& -\frac{1}{\tau_{\Phi}} C_0 (\Phi - \Phi_0) dt -n d\xi \label{nh}\\
dn &=& -\frac{1}{\tau_{\Phi}} C_0 (\Phi - \Phi_0) dt + \tilde{n} dW_t, \label{langevin}
\end{eqnarray}
it becomes clear that Eq. \ref{nh} approximates the spectrum required by the FDT with a single discrete frequency. For poorly ergodic systems, as is the case here, it is generally recognized that a single frequency is insufficient. Using a Nose-Hoover chain as proposed by Bonnet \emph{et al.} \cite{otani}, the spectrum is then approximated by a set of discrete frequencies. For water in particular, which has a rather discrete spectrum itself with large phonon gaps, it is advantageous to utilize a continuous spectrum. By design, Eq. \ref{langevin} satisfies the required spectrum exactly. Compared to the reversible integrator proposed by Martyna \emph{et al.} \cite{m1,m2} or a self-consistent solution of the Nose-Hoover equations of motion, Eq. \ref{langevin} is straightforward to integrate since it has an analytical propagator, cf. next section.

\section{Practical Implementation}

The practical implementation of our scheme is extraordinarily easy: compared to any standard molecular dynamics code, there is only a single additional equation of motion, namely Eq. \ref{dpe}. The Eqs. \ref{dpi} and \ref{dqi} can be solved by any standard discrete time step MD scheme, e.g., using the velocity verlet algorithm. Eq. \ref{dpe} is a stochastic differential equation of a specific type, the Ornstein-Uhlenbeck process:
\begin{equation}
dx = -\frac{1}{\tau}x\ dt + \sqrt{D}\ dW_t.
\end{equation}
The Ornstein-Uhlenbeck process can be integrated analytically to a finite time step \cite{gillespie2}:
\begin{equation}
x(t + \Delta t) = x(t) e^{-\frac{\Delta t}{\tau}} + N \sqrt{\frac{D \tau}{2} \left( 1 - e^{-\frac{2 \Delta t}{\tau}} \right)}. \label{OUupdate}
\end{equation}
$N$ is a Gaussian random number with $\langle N \rangle = 0$ and $\langle N^2 \rangle = 1$. With $x := \Phi - \Phi_0$, $\tau := RC_0$, $D := k_B T / C_0$, we use Eqs. \ref{fdt2} and \ref{OUupdate} to integrate Eq. \ref{dpe} to:
\begin{align}
n(t + \Delta t) = n(t)\ -& \ C_0 \left(\Phi(t) - \Phi_0 \right) \left(1-e^{-\frac{\Delta t}{\tau_{\Phi}}} \right) \nonumber\\ \ +& \ N \sqrt{k_B TC_0 \left(1 - e^{-\frac{2 \Delta t}{\tau_{\Phi}}}\right)}.  \tag{7} \label{ft}
\end{align}

As mentioned above, the implementation of our approach is simple. For example, we performed a thermopotentiostat simulation using only the internal scripting language of the LAMMPS \cite{lammps} molecular dynamics package. We provide a complete thermopotentiostat example simulation for TIP3P water sandwiched between two electrodes as part of the supplementary information. The provided input file "control.inp" can be executed by the standard LAMMPS distribution without any modifications to the LAMMPS code at all.

\section{Choice of $C_0$}

$C_0$ is the geometric capacitance of the two electrode system in absence of the dielectric. It can be estimated using the classical electrostatic expression $C_0 = \epsilon_0 A / d$. Alternatively, it can be measured by an explicit MD simulation for the bare electrodes without the dielectric using $C_0 = \langle n \rangle / \langle \Phi \rangle$. It then enters Eq. \ref{ft} as a parameter and is held constant during the simulation.\\

\section{Thermopotentiostat MD with an explicit Thermostat}

The approach presented here does not require an additional thermostat, since the coupling to the electric field fluctuations is sufficient to thermostat the total kinetic energy of the system. Still, for computational efficiency an explicit thermostat may be desirable, e.g. to achieve fast equilibration. Introducing an explicit thermostat amounts to choosing a suitable expression for $g_i$ in Eq. \ref{dpi}. In standard Langevin dynamics $g_i$ is given by \cite{bussi2}
\begin{equation}
g_i dt = -\frac{1}{\tau} p_i dt + \sqrt{\frac{2}{\tau} k_B T m_i} dW_i.
\end{equation}
Note the conceptual similarity to Eq. \ref{dpe}, where the capacitance $C_0$ plays the role of the mass. For the BDP thermostat, a global $g$ is used which is the same for every kinetic degree of freedom $i$ \cite{bussi2}:
\begin{equation}
g dt = \frac{1}{2 \tau} \left[ \left( 1 - \frac{1}{N_f} \frac{\bar{K}}{K} - 1 \right) \right] p_i dt + \sqrt{ \frac{\bar{K}}{\tau N_f K}} p_i dW.
\end{equation}
Here, $N_f$ is the number of kinetic degrees of freedom and $K$ ($\bar{K}$) is the instantaneous (mean) kinetic energy, respectively. Furthermore, we expect also the Lowe-Andersen thermostat \cite{lowe} to work well in conjunction with our thermopotentiostat approach.

\section{Numerical Details}

Water is described by the TIP3P \cite{tip3p} potential. The electrodes are modeled as single sheets of fixed atoms, where interactions with water are described by the TIP3P O-O pair potential. We do not include any pair potential for the electrode atoms and the hydrogen atoms of the water molecules. The electrode thereby affects the orientation of interfacial water only by the reduced dimensionality at the interface and by the electrode charge. Periodic boundary conditions are applied in $x$ and $y$ direction; non-periodic boundary conditions are used along the direction of the electrode surface normal. Simulations were carried out either in the NVE ensemble or the NVT \cite{bussi1} ensemble at $T = 350$ K, using a time step of 40 atomic units ($\approx 0.97$ fs). All ensembles were first equilibrated without a field for 1 ns using the Langevin thermostat and subsequently for another 2 ns using the BDP thermostat. Sampling time was 4 ns. Coulomb long-range interactions were treated by a particle-particle particle mesh solver \cite{hockney} with a precision of $10^{-5}$. All calculations were performed for rigid water in order to enable the use of a longer time step. Bond lengths and angles were constrained using the SHAKE algorithm \cite{ryckaert,andersen2}.

The shape of the spectral density shown in Fig. \ref{spect} motivates the choice of the thermopotentiostat's relaxation time constant $\tau_{\Phi}$. A smaller $\tau_{\Phi}$ implies a wider white noise region. Decreasing the relaxation time thereby leads to a faster sampling of the phase space, but at the same time disturbs the dynamics of the trajectory more. In practice, we chose a relaxation time of $\tau_{\Phi} = 100$ fs. This value is large enough to ensure that the vibrational density of states obtained from our TIP3P thermopotentiostat MD simulations is virtually indistinguishable from a pure NVE simulation. Simultaneously, the required sampling time to obtain converged dielectric properties is still in range of first principles simulations.

\section{Calculation of Dielectric Profiles}

In analogy to the constant charge simulations in Ref. \cite{loche}, we calculate first the electric field profile $\Delta E_{\perp}(z)$ induced by our thermopotentiostat's displacement field $D_{\perp}$:
\begin{equation}
\Delta E_{\perp}(z) = \epsilon_0^{-1} [D_{\perp} - m_{\perp}(z) + m_{\perp,0}(z)]. \label{DeltaE}
\end{equation}
Subscript 0 denotes zero external electric field. The local polarization density $m_{\perp}(z)$ is obtained from the dielectric's charge density $\rho$ with
\begin{equation}
m_{\perp}(z) = - \int_0^z \rho(z')dz'.
\end{equation}
The local dielectric profile is then extracted from the linear response relation $\Delta E_{\perp} \approx [\epsilon_0 \epsilon_{\perp}]^{-1} D_{\perp}$.

It is instructive to use our theoretical values for the dielectric constant $\epsilon_{gap}$ and $\epsilon_{bulk}$ to estimate $\epsilon_{if}$ for interfacial water from Fumagalli \emph{et al.}'s measurements. Assuming a hydrophobic gap of constant size with \mbox{$d_{gap} = 2$ \AA}, the optimum fit is obtained for $\epsilon_{if} = 2.85$, cf. pink solid line in Fig. 3a, which is significantly lower than what we obtain from our thermopotentiostat simulations. Alternatively, assuming $\epsilon_{if} = 17.3$ for interfacial water to be independent of the specific electrode surface, a fit with identical accuracy is obtained for $d_{gap} = 4$ \AA.

\bibliographystyle{apsrev4-1}